\documentclass[usenatbib,usegraphicx]{mn2e}
\usepackage{bm}
\usepackage{multicol}
\usepackage{txfonts}
\usepackage{times}

\fontencoding{T1}

\newcommand{\WMAP} {WMAP~} 
\newcommand{\be}{\begin{equation}}
\newcommand{\ee}{\end{equation}}
\newcommand{\bea}{\begin{eqnarray}}
\newcommand{\eea}{\end{eqnarray}}
\newcommand{\bold}{\mathbf}
\newcommand{\bhat}{\hat \mathbf}

\title{The effect of reionisation on the CMB--density correlation}

\author[T.~Giannantonio~\&~R.~Crittenden]{Tommaso Giannantonio\thanks{E-mail: tommaso.giannantonio@port.ac.uk} and Robert Crittenden\thanks{E-mail: robert.crittenden@port.ac.uk}\\
Institute of Cosmology and Gravitation, University
of Portsmouth, Mercantile House, Hampshire Terrace, Portsmouth, Hampshire, PO1 2EG, UK}

\begin{document}

%\date{Submitted: 27 March 2007}
\pagerange{\pageref{firstpage}--\pageref{lastpage}} \pubyear{2007}

\maketitle

\label{firstpage}

\begin{abstract}
In this paper we show how the rescattering of CMB photons after cosmic reionisation can give a significant linear contribution to the temperature--matter cross-correlation measurements.
These anisotropies, which arise via a late time Doppler effect, are on scales much larger than the typical scale of non-linear effects at reionisation; they can contribute to degree scale cross-correlations and could affect the interpretation of similar correlations resulting from the integrated Sachs--Wolfe effect.  While expected to be small at low redshifts, these correlations can be large given a probe of the density at high redshift, and so could be a useful probe of the cosmic reionisation history.   
\end{abstract}

\begin{keywords}
cosmic microwave background --- large-scale structure of universe.
\end{keywords}

\section{Introduction}

Power spectrum measurements of the CMB temperature and the large scale distribution of matter have long been important tools for probing cosmology, but only recently has the study of correlations between these important probes been proved to be as useful. Indeed, the absence of strong cross-correlations was instrumental in showing that the CMB anisotropies we observe were of cosmological origin \citep{Bennett:1993kk}. However, the relatively small cross-correlations that do exist give us an important way of constraining those subdominant temperature anisotropies that are created relatively locally.  

Two of the most important local sources of temperature anisotropies contributing to the cross-correlations are the integrated Sachs--Wolfe (ISW) effect \citep{Sachs:1967er} and the Sunyaev--Zeldovidich (SZ) effect \citep{Sunyaev:1970eu}.  Both of these effects provide means of following the growth rate of structure at late times, and through it can tell us about the recent accelerated expansion of the Universe.  The ISW fluctuations are generated when the gravitational potential begins to evolve at late times due to the influence of dark energy; this is a linear effect and the expected cross-correlations are on large angular scales \citep{Crittenden:1995ak}.  On the other hand, SZ fluctuations arise from scattering from hot ionised electrons in clusters, and so lead to smaller angular cross-correlations. 

Recently, evidence for both of these effects has been seen in cross-correlation studies between the CMB anisotropies seen by the WMAP satellite \citep{Bennett:2003bz,Hinshaw:2006ia} and various surveys of large scale structure.  Large angular correlations consistent with the ISW effect have been observed in cross-correlations with radio, infrared, x-ray and optical data, particularly that from the Sloan Digital 
Sky Survey (SDSS) \citep{Boughn:2003yz, Nolta:2003uy, Afshordi:2003xu, Fosalba:2003iy, Scranton:2003in, Fosalba:2003ge, Padmanabhan:2004fy, Cabre:2006qm, Rassat:2006kq, Giannantonio:2006du}.   The SZ effect is usually observed in targeted cluster observations, but evidence for it has also been seen in cross-correlations studies, for example between WMAP and the 2MASS infrared survey \citep{Afshordi:2003xu}.

While the ISW and SZ effects appear to be the dominant sources of cross-correlations (once more mundane foregrounds have been excluded, e.g. \citet {Giannantonio:2006du}), there are other local sources of correlations which are potentially important to understand in order to accurately interpret the observations.  On large scales, it has recently been shown that gravitational magnification can project local inhomogeneities onto higher redshift surveys, causing the matter density at those redshifts to appear more correlated with the ISW anisotropies than would be expected otherwise \citep{LoVerde:2006cj}. On smaller scales, effects like the kinetic SZ, the closely related Ostriker--Vishniac effect \citep{OsVis:1986} and the Rees-Sciama effect \citep{Rees:1968} could tell us much about the evolution of structure, and particularly help probe its velocity on these scales \citep{Iliev:2006zz, McQuinn:2005ce, Schaefer:2005up, Stebbins:2006tv}.

In this paper, we examine another possible source of cross-correlations. Like the kinetic SZ and Ostriker--Vishniac effects, it results from Doppler scattering off of moving electrons, leading to CMB anisotropies with the same frequency dependence as the primordial anisotropies.  Here we focus on the Doppler anisotropies resulting from the electron velocities arising at linear order, which can appear on large angular scales; these could potentially affect the interpretation of the ISW effect.   The impact of this effect on the CMB temperature power spectrum is well understood, where it is known to be subdominant \citep{Dodelson:1993xz,Sugiyama:1993dq,Hu:1993tc,Cooray:1999kg}.  \citet{Alvarez:2005sa} showed that reionisation can produce a significant correlation between the 21cm HI radiation and the CMB; here we focus on its impact on CMB-galaxy cross-correlation measurements, where the effect can be comparable to the ISW at high redshifts, and if unaccounted for, would bias the estimation of parameters.

In section 2, we discuss reionisation and outline the late time linear contributions to the CMB auto- and cross-correlation spectra; in section 3 we present the results coming from a version of CMBFAST \citep{Seljak:1996is}, an extension of modifications made in \citet{Corasaniti:2005pq}.  We discuss the prospects for observing the effect in section 4, before drawing conclusions in section 5.

\section{The effects of reionisation}
\subsection{Reionisation history} 

Cosmic reionisation is currently thought to be caused by the UV radiation emitted by the first luminous objects, and is experimentally constrained by the optical depth to electron scattering of CMB photons, given by \WMAP as $ \kappa_r = 0.092 \pm 0.030 $ \citep{Spergel:2006hy}.  It is also constrained by measurements of the Gunn--Peterson troughs in the Lyman-$\alpha$ part of the spectra of distant quasars from the SDSS \citep{Fan:2001vx,White:2003sc}.   These suggest that the inter-galactic medium (IGM) was fully ionised out to a redshift $ z_r'~=~6.10~\pm~0.15 $ \citep{Gnedin:2006uz}.  As noted by \citet{Shull:2007az}, the scattering up to this redshift accounts for nearly half of the total observed optical depth.  

\begin{figure} 
\begin{center} 
\includegraphics[angle=0,width=0.9\linewidth]{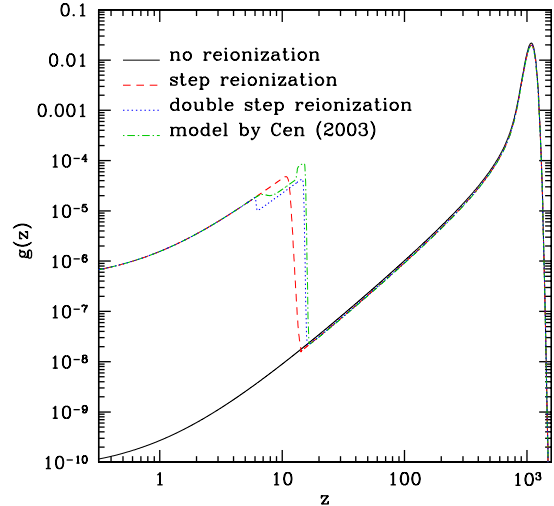}
\caption{Visibility function $ g (z) $ for the best fit \WMAP third year $ \Lambda $CDM model, with and without reionisation. Double reionisation models are also shown, for a double step case and for the \citet{Cen:2002zc} scenario.}
\label{fig:visf}
\end{center}
\end{figure}

Precisely how reionisation happened prior to this is a topic of some debate; the minimal assumption is that the universe became completely ionised very quickly at a single redshift (around a redshift of $z_{r} = 11$ to be consistent with the optical depth constraint.)  However, the process could have been more complex. In the following, we focus on two other models: a ``double step'' model with ionisation fraction brought first to 1/2 at $ z_1 = 15 $ and then to 1 at $ z_2 = 6 $, and a parametrisation of the double reionisation model by \citet{Cen:2002zc}, which again has two distinct phases at  $ z_1 = 15 $ and $ z_2 = 6 $, but with an evolving ionisation fraction between them.
Figure \ref {fig:visf} shows the visibility function $g(z)$ as a function of redshift in these different reionisation scenarios.   Here, the visibility function gives the probability that a photon last scattered at a given redshift, and is related to the optical depth by $ g(r) \equiv n_e \sigma_T a e^{-\kappa(z)} $.

\subsection{CMB anisotropies from reionisation} 

When reionisation is introduced, a second peak appears in the visibility function, corresponding to the restored coupling between photons and matter at late times. Because the CMB photons can now again scatter off free electrons, their properties will thus be altered by the temperature, potential, and velocity of the scatterer, exactly in the same way it happened at the last scattering surface at $ z = 1100 $;  these however are suppressed by the small fraction of photons that are rescattered at this time. 
At linear order, there are three main effects of reionisation \citep{Sugiyama:1993dq, Hu:2001bc}:
\begin {itemize}
\item {Damping of anisotropies on small scales,}
\item {Secondary anisotropy production after reionisation,}
\item {Additional polarisation on large scales.}
\end {itemize}
In addition, other nonlinear effects such as the Ostriker--Vishniac effect and patchy reionisation will be relevant at smaller angular scales ($\ell > 1000$).
The small scale damping is by far the most dominant effect, affecting the temperature power spectrum at small scales ($ \ell \gg 10 $) by a factor $ \sim e^{-\kappa} $.  The polarisation production is also well studied and is responsible for breaking the degeneracy between the optical depth and the spectral index of the primordial perturbations in the WMAP data \citep{Page:2006hz}. 

Here we focus on the impact of the anisotropies generated after the photons rescatter, which are largely dominated by the motion of the scattering electrons.  Along any given line of sight, one expects electron velocities to be moving towards us at some redshifts and moving away from us at other redshifts,
leading to a cancellation of the Doppler effects.  However, the scattering probability is not uniform, causing the Doppler effect to be dominated by the redshifts soon after reionisation.  For this reason,
there is a net anisotropy produced in the CMB temperature, which is correlated with the additional polarisation produced that this time.  The greater the gradient of the scattering probability, the less the cancellation and the larger the anisotropies.  

\begin{figure}
\begin{center} 
\includegraphics[angle=0,width=0.9\linewidth]{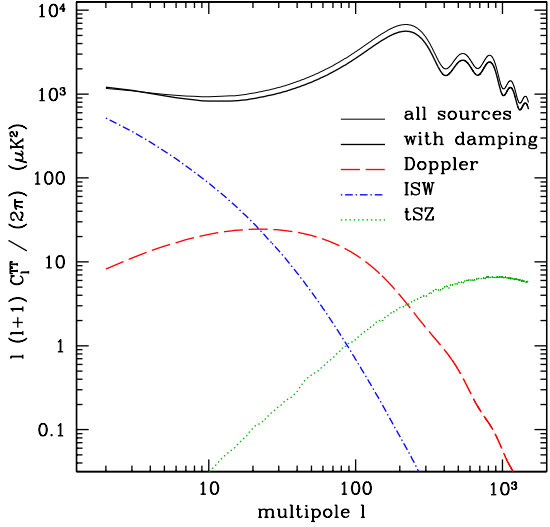}
\caption{Total temperature anisotropies power spectrum for the best fit \WMAP third year $ \Lambda $CDM model, with (thick solid line) and without (thin solid line) reionisation.  The Doppler anisotropies produced after reionisation (dashed line) are significantly smaller than the ISW anisotropies (dot-dashed line) on very large scales, but can be comparable on degree scales. At even smaller scales, the thermal SZ effect becomes relevant (the dotted theoretical curve is obtained by \citet{Schafer:2004zw} using the $N$-body simulation method by \citet{Springel:2001qb}), although it can be distinguished from the other effects taking advantage of its frequency dependence. The reionisation model is the default single step.}\label{fig:allTT}
\end{center}
\end{figure}

Figure \ref{fig:allTT} shows the total anisotropy power spectrum without and with reionisation and the single late time contributions in the reionisation case. The damping on small scales is the biggest contribution, and we can see the small role played by ISW and velocities, while the combined effect of density and gravitational potential perturbations is negligible, as found by \citet{Hu:1993tc} and \citet{Dodelson:1993xz}. In a scenario without reionisation, gravitational effects like the ISW would be the only source of anisotropies at late times.  The Doppler contribution, while always subdominant, is actually larger than the ISW contribution on sufficiently small scales.  At even smaller scales, the thermal SZ effect is important, although it can be distinguished from the other effects taking advantage of its frequency dependence.

More interesting is to consider how the Doppler contribution might be correlated with the density. 
Consider matter falling into a large over-density at some redshift $z_0$; the matter on the far side ($z > z_0$) will be travelling towards us, while the matter on the near side ($z < z_0$) will be travelling away from us.  Scattering from both sides will contribute to temperature anisotropies, but because the scattering is more likely at higher redshifts, the electrons moving towards us will be more likely to
scatter.  Thus the over-density will be associated with a temperature hot-spot, and in the same way under-densities will be associated with CMB cold-spots. 

In the following, we will show that the effect of this secondary rescattering, though always negligible for the temperature power spectrum, is present and can be important in the temperature--matter correlations at high redshift, and in particular is comparable to the magnification bias effect; therefore this effect must be taken into account to produce precise cross-correlation predictions.

\subsection {Power spectra}

We next describe how to calculate the power spectra for these secondary Doppler temperature anisotropies and their cross-correlations with the density.   
These will be proportional to the fraction of photons that are scattered and to the velocity of the scattering electrons $ \textbf{v} $ relative to the observers line of sight $\bhat n$:
\bea
\Delta^T (\bhat n)  = - \int_{\tau_i}^{\tau_0} d \tau  \, g(\tau) \, \bold v (\bold{x}, \tau) \cdot \bhat n ,  
\eea
where $\bold{x} = (\tau_0 - \tau) \bhat{n}$, $\tau_0$ is the present conformal time and $\tau_i$ is taken to be the time just prior to the beginning of reionisation. 
While the physical effect will be gauge invariant, this expression is implicitly in the Newtonian gauge. 
By the epoch of reionisation, the electrons will have fallen into the dark matter wells and will be effectively at rest with respect to the dark matter particles; in the synchronous gauge, where the numerics are usually performed, their peculiar velocities will be very small.  

\begin{figure}
\begin{center} 
\includegraphics[angle=0,width=0.9\linewidth]{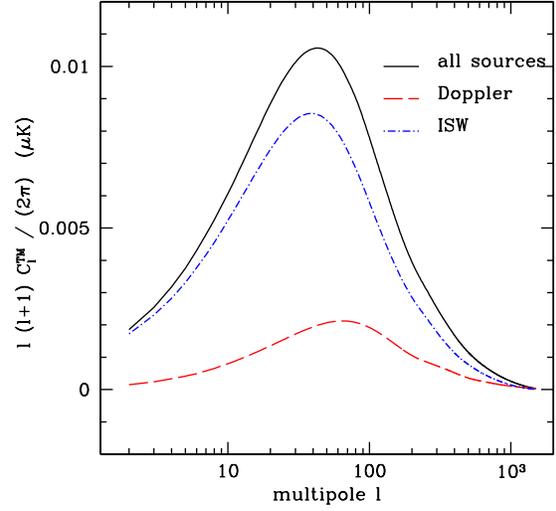}
\caption{Total temperature--density correlation power spectrum for the best fit \WMAP third year $ \Lambda $CDM model and a matter visibility function centred in $ \bar z = 3 $, with reionisation. The reionisation model is the single step and the galaxy bias is set to 1.}\label{fig:allTM}
\end{center}
\end{figure}

\begin{figure*}
\begin{center} 
\includegraphics[angle=0,width=\linewidth]{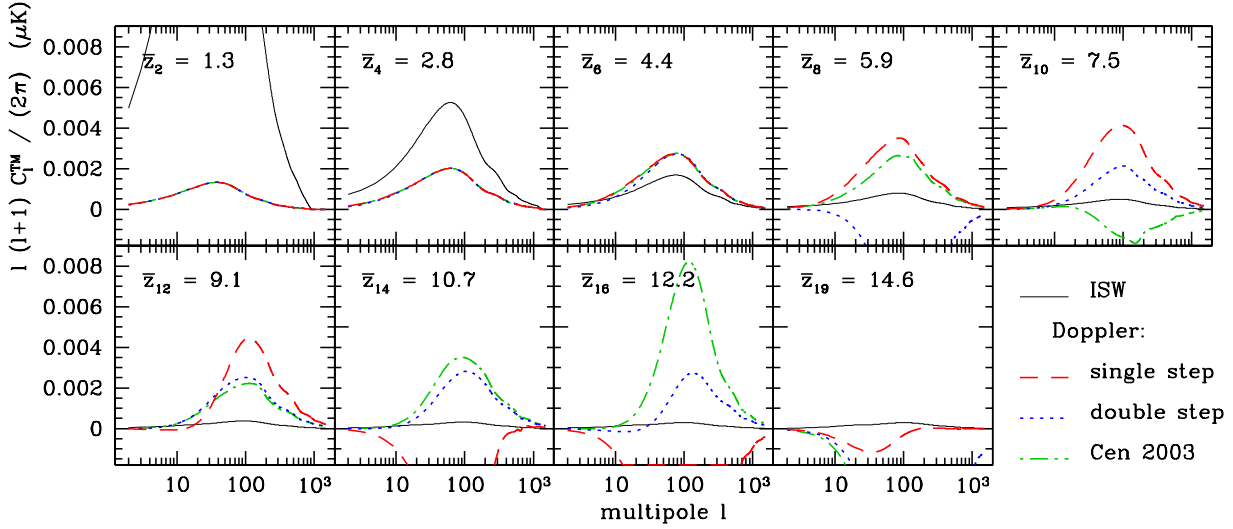}
\caption{Temperature--density correlation power spectrum for the best fit \WMAP third year $ \Lambda $CDM model in function of $ z $, with the set of galaxy selection functions defined in Eq. (\ref {eq:win3}) (redshift tomography). The Doppler effect is compared with the ISW. We see the different evolution of the three different reionisation histories described in Figure \ref {fig:visf}: to every change in the sign of the visibility function derivative corresponds a change in the sign of the cross-correlation due to the Doppler effect. The galaxy bias is always set to 1.}\label{fig:allTMz}
\end{center}
\end{figure*}

We usually work in the harmonic space using a Fourier expansion for quantities like the velocity
\be
\bold v (\bold{x}, \tau) = \sum_{\bold k} \bhat{k} \, v (\bold k, \tau) \, e^{i \bold k \cdot \bold x},
\ee
(where we have assumed the velocity is irrotational.)   Following \citet{Ma:1995ey}, we can relate the Newtonian gauge velocities to those in the synchronous gauge by, 
\be
\bold v = \bold v_s - i \bold k \alpha,
\ee
where $ \alpha \equiv \frac {\dot h + 6 \dot \eta} {2 k^2} $, $h$ and $\eta$ parameterise the metric degrees of freedom in synchronous gauge and dots refer to derivatives with respect to the conformal time. Since we know the synchronous gauge velocities will be small ($v_s \simeq 0$), the Doppler term will be dominated by the term proportional to $\alpha$.   We can relate this to the dark matter density fluctuation using its conservation equation: 
\be
\dot \delta = - \frac {\dot h} {2} = - k^2 \alpha + 3 \dot \eta \simeq - k^2 \alpha.
\ee
The Einstein equations give $ \dot \eta \propto v_s $; so, assuming we can neglect the synchronous velocities, we find:
\be
\bold v \simeq \frac {i \bold k} {k^2} \dot \delta.
\ee

Going back to the anisotropies, and defining as usual $ \mu \equiv {\bhat{k} \cdot \bhat n}$, we have
\be
\Delta^T (\bhat n) = - \int_{\tau_i}^{\tau_0} d \tau  \sum_{\bold k} e^{i k \mu (\tau_0 - \tau)} i \mu g(\tau) \frac {\dot \delta (\bold k, \tau)} {k},
\ee
which can be integrated by parts; dropping the surface terms where the visibility function is small, we find
\be
\Delta^T (\bhat n) = - \int_{\tau_i}^{\tau_0} d \tau  \sum_{\bold k} e^{i k \mu (\tau_0 - \tau)} \frac{1}{k^2} \frac {d} {d\tau} \left[ {g(\tau) \dot \delta (\bold k, \tau)} \right].
\ee
If we assume linear theory for the growth of density perturbation so that $ \delta (\bold k, \tau) = D (\tau) \delta (\bold k, 0) $, we can then expand the exponential in terms of spherical harmonics to show
\bea
\Delta^T (\bhat n) &  = & - \int_{\tau_i}^{\tau_0} d \tau \sum_{\bold k} \delta (\bold k, 0) \frac{4 \pi}{k^2} \sum_{l, m} i^l j_l (k(\tau_0 - \tau)) Y_{lm} (\bhat n) Y^*_{lm} (\bhat k) \nonumber \\
& \times & \frac {d} {d\tau} \left[ {g(\tau) \dot D (\tau)} \right].
\eea
From this expression, we find the harmonic coefficients for the Doppler anisotropies to be 
\be
a^{T}_{lm} = \sum_{\bold k} \delta (\bold k, 0) \, 4 \pi i^l \, Y^*_{lm} (\bhat k) \, \Delta_T^l(k),
\ee
where
\be
\Delta^T_l(k) = - \frac{1}{k^2} \int_{\tau_i}^{\tau_0} d \tau \, j_l (k(\tau_0 - \tau)) \, \frac {d} {d\tau} \left[ {g(\tau) \dot D (\tau)} \right] .
\ee
The expectation of the square of these coefficients gives the Doppler power spectrum
\be \label {eq:clbasic}
C_l^{TT} = \frac {2} {\pi} \int dk k^2 P(k) |\Delta^T_l|^2,
\ee
where $ P (k) $ is the matter power spectrum today.

Following similar arguments, we can find similar equations for the anisotropies in the matter density (see e.g. \citet{Boughn:1997vs}). Specifically, we find 
\be
a^{g}_{lm} = \sum_{\bold k} \delta (\bold k, 0) \, 4 \pi i^l \, Y^*_{lm} (\bhat k) \, \Delta_g^l(k),
\ee
but instead 
\be
\Delta_l^g (k) = \int_{\tau_i}^{\tau_0} d \tau b_g(\tau) \, D(\tau) \, W(\tau) \, j_l (k (\tau_0- \tau_i),
\ee
where $ b_g(\tau) $ is the possibly evolving galactic bias and $ W(\tau) $ the matter selection function normalised to unity.
We can then use these to derive the matter auto-correlation and the matter--temperature power spectra
\bea
C_l^{gg} & = & \frac {2} {\pi} \int P(k) k^2  |\Delta_l^g (k)|^2 dk \nonumber \\
C_l^{Tg} & = & \frac {2} {\pi} \int P(k) k^2  \Delta_l^T (k)  \Delta_l^g (k) dk.
\eea
For the following discussion, we implement these in a numerical Boltzmann code, a modified version of CMBFAST.

At sufficiently small angular scales, it is possible to simplify the calculations using the small angle approximation.   Here, we can integrate the complete evolution Eq. (\ref{eq:clbasic}) in an approximate analytical form \citep {Limber54}:
\bea \label {eq:clvel}
C_l^{TT} & \simeq & \frac {2} {\pi} \int \frac {d r} {r^2} P \left( \frac {l + 1/2} {r} \right) \left\{ \frac {d} {dr} \left[-\frac {g(r) \dot D (r)} {(l + 1/2)^2} \right] \right\}^2  \\
C_l^{Tg} & \simeq & \frac {2} {\pi} \int \frac {d r} {r^2} P \left( \frac {l + 1/2} {r} \right) b_g W(r) D(r) \frac {d} {dr} \left[ -\frac {g(r) \dot D (r)}  {(l + 1/2)^2} \right] , \nonumber
\eea 
where $r \equiv \tau_0 - \tau$ is the conformal distance. 
However, on large scales this approximation will not be valid, and therefore throughout we compute all our results using the full-sky Boltzmann code.

We can see from Eq. (\ref {eq:clvel}) how this effect depends on the cosmological parameters: the derivative of the visibility function $ g $ will bring a dependency on the reionisation history, as we will see in more detail in section \ref {sec:results}; on the other hand, the derivative of the growth factor depends on the matter parameters $ \Omega_m, \sigma_8 $, and also on the dark energy equation of state. However, in typical models the visibility function will change more dramatically than the growth function, and will dominate the effect.

Here we have focused on the cross-correlation with the matter density;  the cross-correlation with the 21 cm line depends on instead on the neutral hydrogen density which can be more complicated \citep{Alvarez:2005sa}.   If reionisation were uniform, peaks in the matter density would be associated with a higher neutral hydrogen density; however, ionisation is expected to be produced by the early structures formed in the highest peaks, causing the peaks in the matter density to have less neutral hydrogen.  Depending on the competition between these effects, the CMB-21 cm cross-correlation can have the opposite sign from what is derived above \citep{Alvarez:2005sa}.  However for the discussion below, we will always assume the tracers are positively correlated with the total matter density.  

\subsection {Double scattering}
As noted by \citet{Cooray:1999kg}, double scattering processes can produce higher order corrections to the Doppler signal. These corrections have a power spectrum which, for scales smaller than the width of the visibility function, can be approximated by
\be
\Delta_l^{T, ds} \simeq \frac {1}{k^2} \int_{\tau_i}^{\tau_0} d \tau j_l [k (\tau_0 - \tau)] g (\tau) \dot \kappa (\tau) \dot D (\tau).
\ee
Their effect is small, always $ < 5 \% $ at the peak, and we will therefore neglect it, this being anyway a conservative choice.

\section {Results} \label {sec:results}
\subsection {Low redshift signal}

The expected cross-correlations will depend on many factors, but the biggest are the depth of the survey and the assumed history of reionisation.  To begin, we look at the amplitude of the effect at low redshifts.  Figure \ref{fig:allTM} shows the expected cross-correlation with a broad selection function at mean redshift $ \bar z = 3 $.  This assumes a selection function of the form,   
\be
W_{t} (z) = \frac {3}{2} \frac {z^2} {z_0^3} e^{-\left(\frac {z} {z_0} \right)^{3/2}}
\ee
where $ \bar {z} =  1.41 z_0 $, which is a good approximation to the distribution of objects in a flux limited survey, without any imposed cut in redshift.  For comparison, we show the ISW cross-correlation for the same survey depth.  For this depth, the effects are actually comparable, though this is in part because the ISW effect peaks around $z \simeq 1$ and is already dying off at these redshifts.  

\subsection {Reionisation history dependence}

The low redshift signal is largely independent of assumptions of the reionisation history, since we believe the Universe to be totally ionised at that time.  However, the signal will increase as we probe higher and higher redshifts and will provide us with a way to probe the visibility function evolution.  
To study the time evolution of this effect, we use narrower selection functions corresponding to the imposition of redshift cuts on a survey; to be specific, we use a redshift slicing proposed by \citet {Hu:2004yd} for an LSST type survey.  There they assumed the total selection function could be divided into many narrower bins at different redshifts.  The $i$-th selection function was assumed to be
\be \label {eq:win3}
W_i (z) \propto W_{t} (z) \times \left\{ \mathrm {erfc} \left[ \frac {(i - 1) \Delta - z} {\sigma(z) \sqrt {2}} \right]- \mathrm{erfc} \left[ \frac {i \Delta - z} {\sigma(z) \sqrt {2}}\right] \right\},
\ee
where $ \Delta = 0.8 $, $ \sigma(z) = 0.02 (1+z) $ and every slice is normalised to unity. To study the evolution out to the time of reionisation, we will consider $ i = 1,...,19 $, corresponding to mean redshifts $ \bar z_i$ from 0.5 to 15.

We can see how the effect evolves with redshift in Figure \ref {fig:allTMz}.  The ISW effect is dominant at low redshifts (where dark energy is dynamically important) and becomes negligible at high redshifts.  The Doppler contribution largely depends on the time derivative of the visibility function.  At late times, once the Universe is fully reionised, the visibility function decreases smoothly as the universe expands and the density of scatterers decreases; the resulting cross-correlations evolve relatively slowly.   However, the picture can change dramatically once we reach redshifts where the ionisation fraction changes, which can drastically alter the visibility function.  This is particularly the case if we had a survey sensitive to the beginning of the ionisation history, which for the \WMAP 3 best fit model is at $ z \simeq 11 $.  There, the sign of the correlation becomes negative and the amplitude can be large if ionisation occurs quickly.

This dependency on the visibility function makes it a sensitive probe of the reionisation history.
In Figure \ref {fig:allTMz}, we compare the signal for different histories: the single step reionisation, a simple double step reionisation, and a more complicated reionisation based on \citet{Cen:2002zc}.  At low redshifts, all models are fully ionised and have equivalent cross-correlations.   When the histories diverge ($z > 4$), the cross-correlation can be reduced or change signs twice if we are able to see two epochs of the visibility function increasing.   

The most prominent feature is actually the negative cross-correlation which occurs at the primary epoch of reionisation.  Assuming a single step process, the amplitude of this effect depends on two factors: the total optical depth and how long it takes to ionise the universe.  The total optical depth determines the redshift of reionisation as well as height of the peak of the visibility function.  However, since the time derivative of the visibility function comes into the integral, the duration of reionisation is also important.

The full impact of these factors is most easily seen in the temperature spectrum, which sums the anisotropies from all redshifts.  
Increasing the optical depth raises the signal and shifts the spectrum to slightly smaller scales.  On the other hand, shortening the duration of reionisation allows the resolution of velocities on smaller scales, and so adds small scale power to the anisotropies.

\subsection {Comparison with magnification}

It is interesting to compare the late Doppler effect with the gravitational magnification effect described by \citet{LoVerde:2006cj};  while no additional temperature fluctuations are created by the magnification, 
this effect can make the low redshift density distribution appear to be in a higher redshift sample, thereby causing it to be more correlated with the ISW anisotropies than one might expect.  The amplitude of these will depend on how the number density of objects depends on the flux cutoff.  Since no new anisotropies are created, the effect introduces covariances between the low and high redshift cross-correlations. 

\begin{figure}
\begin{center} 
\includegraphics[angle=0,width=0.9\linewidth]{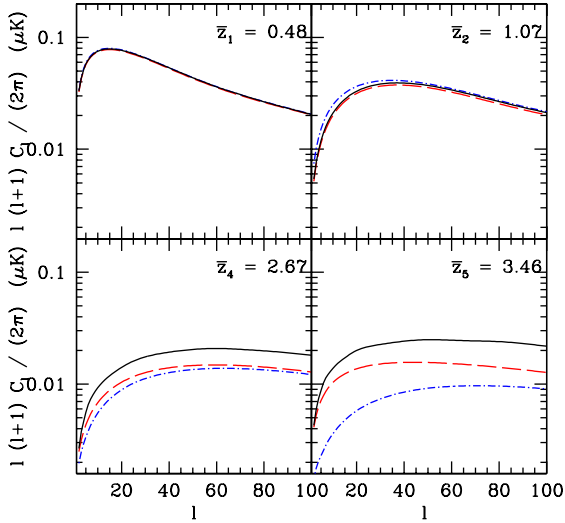}
\caption{Comparison of the magnification and Doppler cross-correlations. We show the ISW effect alone (blue, dot-dashed), the effect of adding corrections from cosmic magnification (red, dashed) and all three effects together (black, solid). 
Here we have reproduced the calculations of \citet{LoVerde:2006cj} using the same cosmology and assumptions about the galaxy samples (mock catalogue I) and $dN/dz$.
}\label{fig:allTMmariI}
\end{center}
\end{figure}

\begin{figure}
\begin{center} 
\includegraphics[angle=0,width=0.9\linewidth]{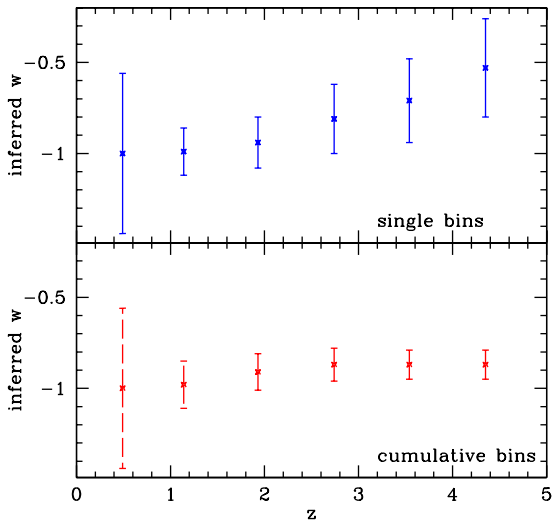}
\caption{The inferred value of the dark energy equation of state parameter $ w $ obtained ignoring the effect of reionisation, for the samples of \citet{LoVerde:2006cj}. The input model had $w = -1$. The top panel shows the result for the single redshift bins, while in the bottom panel we present the effect of adding up to the $i$-th bin.}
\label{fig:infer6}
\end{center}
\end{figure}

\citet{LoVerde:2006cj} showed that ignoring the magnification effect can bias the value of the dark energy equation of state which would be inferred from the cross-correlation measurements.  We show 
in Figure \ref{fig:allTMmariI} that the additional cross-correlations from the Doppler effect are of a similar magnitude to the gravitational magnification effect, so one would expect a similar kind of bias in the equation of state if they were ignored.  But unlike the magnification effect, the Doppler correlations represent true new correlations at these redshifts, so the covariance between the higher and lower redshift measurements will be much smaller.
We studied the bias that would be obtained by ignoring the Doppler correlation (but including magnification) with a Fisher matrix analysis, as done by \citet{LoVerde:2006cj}. The inferred values for $ w $ obtained with marginalising over the other parameters $\Omega_m$, $\Omega_b$, $n_s$, $h$ and $\sigma_8$ with priors errors of $5\%$ are shown in Figure \ref {fig:infer6}. The overall inferred $w$ differs from the ``true'' one that we put into the calculation by almost $2 \sigma$.

\section {Observability}

The ISW effect itself is difficult to detect from cross-correlation measurements, principally because of the presence of large primordial CMB anisotropies originating at a redshift of $z \sim 1000$.  These fundamentally limit the significance which the the cross-correlations can be observed to a signal to noise level of $7-10$.  It is worth exploring if the late Doppler anisotropies are similarly limited.     

\begin{figure}
\begin{center} 
\includegraphics[angle=0,width=0.9\linewidth]{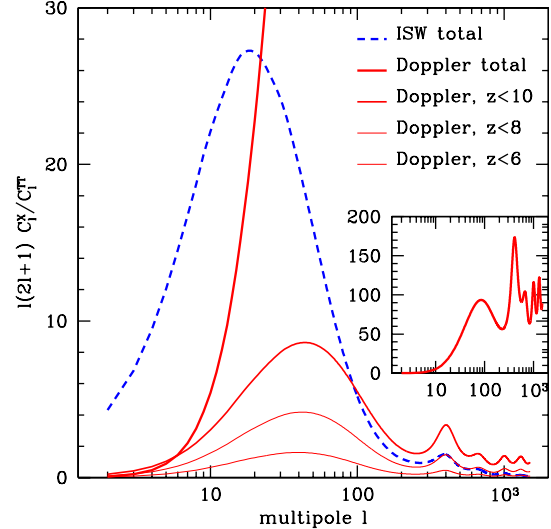}
\caption{The total signal to noise for ISW and rescattering Doppler
effect, for the single step reionisation history. The different lines represents different redshift cuts, respectively from top to bottom: no cut and cuts at redshift 10, 8, 6. The insert shows the total signal to noise (without redshift cuts) for the velocity effect.  At high $\ell$, cross-correlations from non-linear effects will become important. }\label{fig:SNRob}
\end{center}
\end{figure}

We first consider the most optimistic picture possible, in which we had a full sky map out to a given redshift and were able to reconstruct a map of the predicted late Doppler anisotropies.  In this case the expected signal to noise is given by \citep{Crittenden:1995ak}
\be
\left( \frac{S}{N} \right)^2 \simeq \sum_{l}  (2 l +1) \frac{C_l^{\mathrm{T-Dop}}}{C_l^{\mathrm{T-tot}}}. 
\ee
By looking at the auto-correlations in Figure \ref {fig:allTT}, we see that for large $\ell $, the Doppler anisotropies exceed the ISW anisotropies.  Thus we might hope that since the signal to noise weights the large multipoles more, the total signal to noise might be higher than the ISW case.  This is indeed the case, as demonstrated by Figure \ref {fig:SNRob}.  When we include the whole signal, which includes the cross-correlations from the epoch of reionisation, the signal to noise can be much larger than for the ISW.  
For a typical model ($ \kappa = 0.092 $) we find a total optimal signal to noise of order $ S/N \sim 20 $.  This can rise to as high as 30 with a higher optical depth and shorter duration of reionisation, or drop as low as 10 if we take these parameters to the other extreme.   However, the figure also shows that the much of the small angle anisotropies arise at very high redshift, and it would take a very deep survey before the significance could exceed more than a few.  

\begin{figure}
\begin{center} 
\includegraphics[angle=0,width=0.9\linewidth]{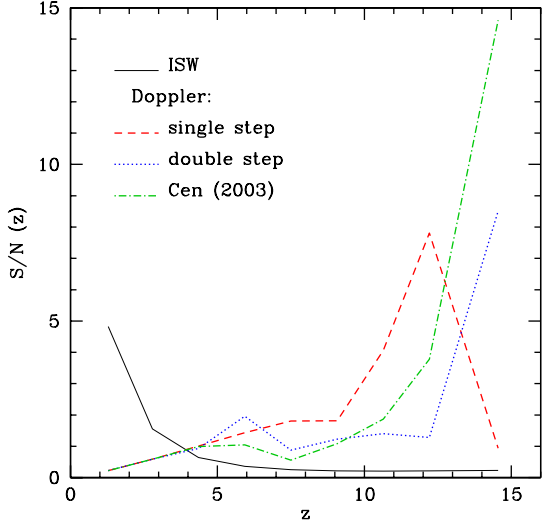}
\caption{Evolution of signal to noise ratio in function of the redshift bins
  for the velocity effect in different reionisation scenarios compared with the ISW.}\label{fig:SNsingle}
\end{center}
\end{figure}

The above calculations are for the most optimistic case.  A more realistic estimate can be found using the calculated cross-correlations functions for the galaxy selection functions described above.  In this case, the signal to noise is given by
\be
\left[ \frac{S}{N} (z_i) \right]^2 = \sum_l (2l+1) \frac { [C^{Tm} (z_i)]^2} {C_l^{mm} (z_i,z_j) C_l^{TT} + [C_l^{Tm}]^2}.
\ee
We show this signal to noise ratio for the Doppler contribution compared to the naive ISW effect in Figure \ref{fig:SNsingle}, for the \WMAP best fit model of quasi instantaneous reionisation and the series of redshift tomography proceeding from the matter visibility functions defined in Eq. (\ref {eq:win3}). We can see how a small but non negligible $S/N$ is produced at high redshift, while a much larger signal is generated at the deeper reionisation epoch.

A small but potentially measurable signal to noise is present at redshifts $ z < 7 $, where we know that we can find galaxies and other collapsed density tracers;  the signal can be enhanced if the ionisation is still evolving at these low redshifts. A much larger signal is expected at higher redshifts, but there the best density probe is probably the 21-cm radiation as suggested by \citet{Alvarez:2005sa}.

\section{Conclusions}

In this paper, we have focused on the linear contributions to cross-correlations arising from the rescattering of CMB photons.  These arise from scales of order $k \sim 0.01 h \, \mathrm{Mpc}^{-1}$ and are seen on degree scales.  However, it should be emphasised that these are by no means the only such sources of cross-correlations; non-linear effects such as patchy reionisation, the Ostriker--Vishniac effect and high redshift SZ sources all could potentially contribute to cross-correlations from the reionisation epoch (see the recent paper by \citet{Slosar:2007sy}).   These cross-correlations will typically be at much smaller angular scales.  While the linear effect is sensitive to the bulk properties of reionisation, such non-linear cross-correlations could potentially provide useful information on the detailed physics of the reionisation process itself. 

At low redshifts, these cross-correlations are small but could still be important in the interpretation of the integrated Sachs--Wolfe signal.  For example, we have shown that they are comparable to the effect of gravitational magnification.  A high redshift ISW signal has also been shown to be a potential means of discriminating dark energy models from those in which the laws of gravity are modified \citep{Song:2006jk}; to test such models correctly, including the cross-correlations from rescattering would be essential.   

Detecting these cross-correlations in their own right will clearly be a challenge since it requires deep probes of the matter density.   One possibility is to use galaxy or quasars found with an optical survey such as the DES (for quasars) \citep{Abbott:2005bi}, LSST \citep{Tyson:2006hs}, and possibly the panoramic survey of the SNAP, which is still in the proposal stage, which could probe the Universe to a redshift of $z=3$.   Alternatively, radio instruments such as LOFAR \citep{Rottgering:2006ms} or the SKA \citep{Blake:2004pb,Torres-Rodriguez:2007mk} could potentially go as deep or deeper by finding 21 cm HI emission from early galaxies; or, if the near-infrared background is dominated by the first generation 
of structure formation \citep{Santos:2002hd,Salvaterra:2002rg}, it might be useful as a proxy for the density at high redshifts.  Finally, as discussed by \citet{Alvarez:2005sa},
the 21-cm emission from the neutral gas at reionisation could be observed directly by an experiment like SKA, providing a probe of the density precisely where the ionisation is changing most dramatically. 

\subsection*{Acknowledgements}

We thank Asantha Cooray, Bob Nichol, Bj{\"o}rn Malte Sch{\"a}fer, Daniel Thomas and Jussi V{\"a}liviita for useful conversations, and Pier-Stefano Corasaniti and Alessandro Melchiorri who collaborated on the initial version of the modified Boltzmann code, that is publicly available at \texttt {http://www.astro.columbia.edu/ \~{}pierste/ISWcode.html}.

\label{lastpage}

\end{document}